\documentclass[a4paper]{aa}
\usepackage{psfig,longtable,lscape}

\usepackage{epsfig}
\def \sax {BeppoSAX}

\def \deg{^\circ}

\def \hcm {\hbox {\ifmmode $ atom cm$^{-2}\else atom cm$^{-2}$\fi}}

\def\approxgt{\mathrel{\hbox{\rlap{\lower.55ex \hbox {$\sim$}}
 \kern-.3em \raise.4ex \hbox{$>$}}}}
\def\approxlt{\mathrel{\hbox{\rlap{\lower.55ex \hbox {$\sim$}}
 \kern-.3em \raise.4ex \hbox{$<$}}}}

\begin{document}


\title{A catalogue of soft X--ray sources in the galactic center region}

\author{L. Sidoli\inst{1},
    T. Belloni\inst{2},
 S. Mereghetti\inst{3}
}

\offprints{L. Sidoli (lsidoli@astro.estec.esa.nl)}

\institute{
       Astrophysics Division, Space Science Department of ESA, ESTEC,
       Postbus 299, NL--2200 AG Noordwijk, The Netherlands
\and
Osservatorio Astronomico di Brera, Via Bianchi 46, I--23807 Merate (Lc),
  Italy
\and
Istituto di Fisica Cosmica ``G.~Occhialini'', CNR, via
  Bassini 15, I--20133 Milano, Italy
}

\date{Received 12 October 2000; Accepted 4 January 2001}

\authorrunning{L. Sidoli et al.}

\titlerunning{{A catalogue of soft X--ray sources 
in the GC region}}

\abstract{We present    
a catalogue of 107 point-like 
X--ray sources
derived from a systematic analysis of all the ROSAT PSPC observations
of the galactic center region performed
in    1992--1993.
Besides  SgrA*, the massive black hole at the galactic center, 
41 X--ray sources have been positionally associated
with already classified objects.
Twenty are identified with foreground stars and 
five with known Low Mass X--ray Binaries.
The majority of the sources in our catalogue 
still remains unidentified. 
They are  hard and/or severely
absorbed and probably represent a large  population
of X-ray binaries located in the galactic 
center region, accreting 
at low accretion rates, and still largely unknown.
} \maketitle

\section{Introduction}
\label{sect:intro}

The galactic center (hereafter GC) region has always been among the privileged
targets of many X--ray missions. The observations made over
the years established that a severe
crowding of X-ray sources exists towards this part of our Galaxy
(see Sidoli et al. (1999) for the results of a  \sax\ survey of the GC 
with its Narrow Field Instruments).
Many of the brightest  sources are probably  Low Mass X--ray Binaries (LMXBs) 
containing
neutron stars or black holes, both with persistent and
transient behavior, while the nature of the faintest sources is unknown.
Many transient sources have  been discovered in the last few years,
especially during the monitoring of the galactic center region with
the  \sax\ Wide Field Camera (Ubertini et al. 1999) 
and the Rossi-XTE All Sky Monitor  (Bradt et al. 2000).

The ROSAT Position Sensitive Proportional Counter (PSPC)
performed a raster scan of the GC region
($|l|<1.5\deg$)$\times$($|b|<2\deg$) in 1992 and 1993.
The  region covered  is rectangular ($3\deg\times4\deg$) with the
major axis  oriented  perpendicular to the galactic plane.

We performed a detailed  spatial analysis of these
data  with the main objective of  obtaining
a  catalogue of X--ray sources in the soft X--ray  energy band.
All the data analyzed here have been retrieved
from the ROSAT public archive in MPE.
These datasets have previously been analyzed by
Predehl \& Tr\"umper (1994)
and by Predehl et al. (1995), but neither of these works
was aimed at the production of a full catalogue of sources in
the region.
 
\section{Observations and Data Reduction}
\label{sect:obs}

The PSPC on board the ROSAT satellite (Pfeffermann et al. 1986)
 covered the energy band 0.1--2.4 keV with a moderate energy resolution
(3--4 energy bands can be defined) and an angular resolution of
$\sim$20$''$ (FWHM).
The detector has a circular field of view of
$\sim$2$\deg$ diameter. The radial and circular supports of the
entrance window  produce some artifacts, whose effects are
mitigated by wobbling the satellite during pointing observations.

The data consist of 43 pointed observations
with exposure times in the range 2000--3000 seconds,
aimed at completely covering the central part of our Galaxy with the inner
region of the PSPC detector.
These observations were performed between 1992 February and 1993 March.
We also included in our sample a single deeper pointing (47,000 sec)
centered on the Sgr~A* position  performed on 1992 March 2.
The log of the observations is reported in Table~\ref{rosat_log}.

 
\begin{table*}[htbp]
\begin{center}
  \caption{The PSPC observations log.}
\label{rosat_log}
    \begin{tabular}[c]{|l|l|l|l|}
\hline 
Observation     & Pointing Direction  & Obs. Start   &  Obs. End \\
 ID  &  R.A. \& Dec. (J2000)       &  dd/mm/yy  hh:mm &  dd/mm/yy  hh:mm \\
\hline 
 rp400150n00  &     17 49 16.80  --30 12 00.0 &  04/03/92 03:48 & 04/03/92 04:28 \\ 
 rp400179n00  &     17 47 57.60  --30 01 48.0 &   02/03/92 19:53 &   02/03/92 20:41  \\
 rp400180n00  &     17 46 38.40  --29 51 00.0 &   02/03/92 08:49 &   02/03/92 09:25 \\
 rp400181n00  &     17 45 19.20  --29 40 48.0 &   02/03/92 01:11 &   02/03/92 03:03  \\
 rp400182n00  &     17 44 00.00  --29 30 36.0 &   01/03/92 20:06 &   01/03/92 20:45 \\
 rp400183a01  &     17 42 43.20  --29 19 48.0 &   28/03/93 19:56 &   28/03/93 20:38 \\
 rp400183n00  &     17 42 43.20  --29 19 48.0 &   02/03/92 11:54 &   02/03/92 12:25 \\
 rp400184n00  &     17 41 24.00  --29 09 00.0 &   01/03/92 10:26 &   10/03/92 12:14 \\
 rp400185n00  &     17 40 07.20  --28 58 48.0 &   29/02/92 21:38 &   29/02/92 22:26 \\
 rp400186n00  &     17 38 50.40  --28 48 00.0 &   29/02/92 15:16 &   29/02/92 19:09 \\
 rp400187n00  &     17 50 04.80  --29 54 36.0 &   03/03/92 16:37 &   03/03/92 10:45  \\
 rp400188n00  &     17 48 43.20  --29 44 24.0 &   03/03/92 11:52 &   03/03/92 17:08 \\
 rp400189n00  &     17 47 26.40  --29 34 12.0 &   02/03/92 21:33 &   02/03/92 22:16 \\
 rp400190n00  &     17 46 07.20  --29 24 00.0 &   03/03/92 08:40 &   03/03/92 09:20  \\
 rp400191n00  &     17 43 31.20  --29 03 00.0 &   04/03/92 19:49 &   04/03/92 20:30  \\
 rp400192n00  &     17 42 12.00  --28 52 12.0 &   01/03/92 12:00 &   01/03/92 17:19  \\
 rp400193n00  &     17 40 55.20  --28 42 00.0 &   01/03/92 03:56 &   01/03/92 04:42 \\
 rp400194n00  &     17 39 38.40  --28 31 12.0 &   29/02/92 20:04 &   29/02/92 20:51 \\
 rp400195n00  &     17 50 50.40  --29 37 48.0 &   03/03/92 19:50 &   03/03/92 20:36 \\
 rp400196n00  &     17 49 31.20  --29 27 00.0 &   04/03/92 07:01 &   04/03/92 07:40 \\
 rp400197n00  &     17 48 12.00  --29 16 48.0 &   03/03/92 18:13 &   04/03/92 13:50 \\
 rp400198n00  &     17 43 00.00  --28 35 24.0 &   02/03/92 10:20 &   03/03/92 15:32  \\
 rp400199n00  &     17 41 43.20  --28 24 36.0 &   01/03/92 07:13 &   01/03/92 07:54 \\
 rp400200n00  &     17 40 26.40  --28 14 24.0 &   29/02/92 23:15 &   01/03/92 00:02 \\
 rp400201n00  &     17 51 38.40  --29 20 24.0 &   03/03/92 23:07 &   03/03/92 23:46 \\
 rp400202n00  &     17 50 19.20  --29 10 12.0 &   04/03/92 08:37 &   04/03/92 09:15  \\
 rp400203n00  &     17 49 00.00  --29 00 00.0 &   04/03/92 18:09 &   10/03/92 15:04 \\
 rp400204n00  &     17 47 40.80  --28 49 48.0 &   04/03/92 11:56 &   07/03/92 10:28 \\
 rp400205n00  &     17 46 24.00  --28 39 00.0 &   02/03/92 18:18 &   03/03/92 13:54 \\
 rp400206n00  &     17 43 48.00  --28 18 00.0 &   01/03/92 18:21 &   02/03/92 15:36 \\
 rp400207n00  &     17 45 55.20  --28 11 24.0 &   01/03/92 13:42 &   02/03/92 13:59 \\
 rp400208n00  &     17 41 14.40  --27 57 00.0 &   01/03/92 04:51 &   01/03/92 06:18 \\
 rp400209n00  &     17 52 24.00  --29 03 00.0 &   04/03/92 05:23 &   04/03/92 06:03 \\
 rp400210n00  &     17 51 04.80  --28 52 48.0 &   04/03/92 02:13 &   04/03/92 02:53 \\
 rp400211a01  &     17 49 48.00  --28 42 36.0 &   23/03/93 09:24 &   23/03/93 09:47 \\
 rp400211n00  &     17 49 48.00  --28 42 36.0 &   03/03/92 21:56 &   03/03/92 22:11 \\
 rp400212n00  &     17 48 28.80  --28 32 24.0 &   16/03/93 21:06 &   16/03/93 21:53 \\
 rp400213n00  &     17 47 12.00  --28 22 12.0 &   16/03/93 22:42 &   16/03/93 23:30 \\
 rp400214n00  &     17 45 55.20  --28 11 24.0 &   02/03/92 07:06 &   02/03/92 07:49 \\
 rp400215n00  &     17 44 36.00  --28 01 12.0 &   01/03/92 21:34 &   01/03/92 22:21 \\
 rp400216n00  &     17 43 19.20  --27 51 00.0 &   01/03/92 23:10 &   01/03/92 23:57 \\
 rp400217n00  &     17 42 02.40  --27 40 12.0 &   01/03/92 08:48 &   01/03/92 09:30 \\
 rp900162n00  &     17 45 40.80  --29 00 00.0 &   02/03/92 03:56 &  09/03/92 05:40 \\ 
 \hline
\end{tabular}
\end{center}
\end{table*}

The data have been analysed using  EXSAS (Extended Scientific
Analysis System, version 98APR;
Zimmermann et al. 1993)
implemented in the ESO--MIDAS version 97NOVpl2.0 on Sun/Solaris.

All the   analysis described below 
was performed   in four different energy ranges: 0.1--2.4 keV
(channels 8--240, total  energy band, T),
0.1--0.4 keV (channels 8--40, soft energy band, S),
0.5--0.9 keV (channels 52--90, medium  energy band, M) and
0.9--2.4 keV (91--240, hard energy band, H).

\begin{figure*}[!ht]
\centerline{\psfig{figure=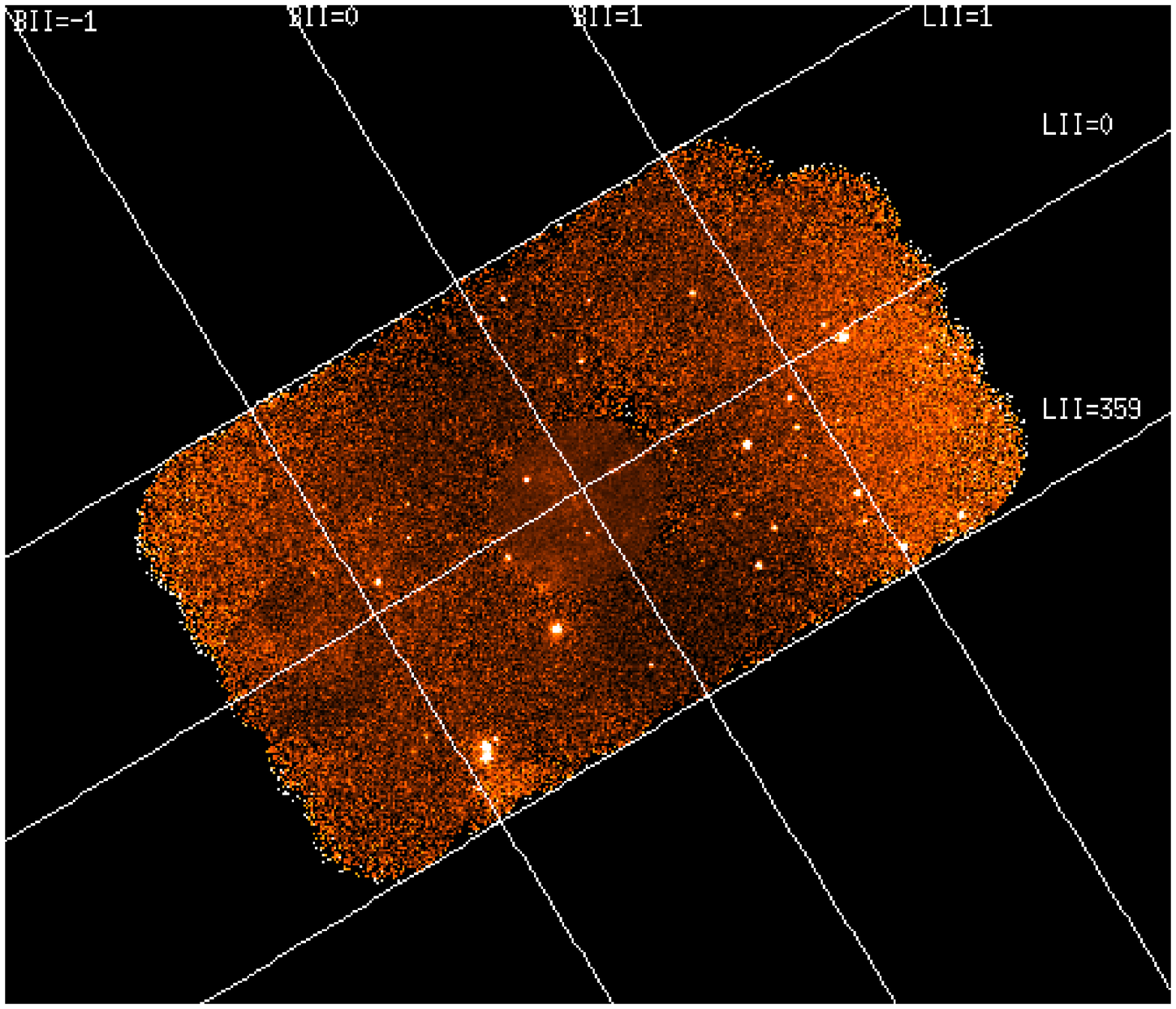,height=110mm,bbllx=70mm,bblly=80mm,bburx=140mm,bbury=250mm}}
\vspace{1cm}
 \caption[]{Mosaic image of the galactic center region in the 0.1-2.4 keV
 energy range. Only the inner part of the PSPC  detector has been used.
The image has been corrected for the exposure and for the vignetting 
}
   \label{fig:totcor}
\end{figure*}

\begin{figure}[!ht]
\vskip 0.5truecm
\centerline{\psfig{figure=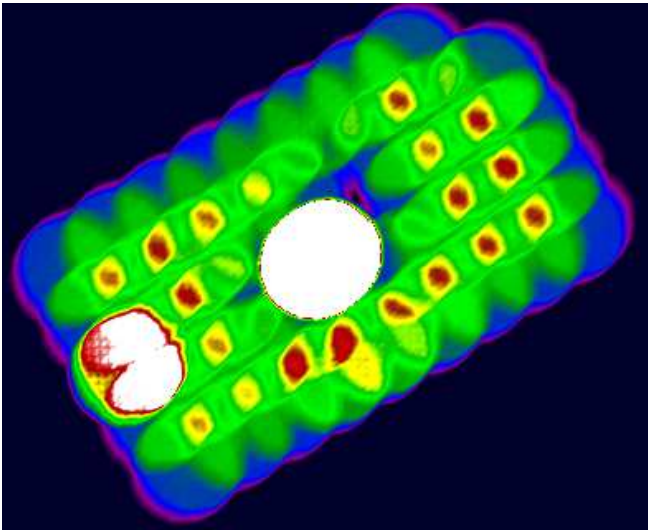,height=70mm,bbllx=86pt,bblly=283pt,bburx=510pt,bbury=629pt,clip=}}
\vskip 0.5truecm
\caption{Exposure map  appropriate for the PSPC mosaic
in the total energy band }
\label{fig:expo}
\end{figure}


\subsection{Source Detection}
\label{sect:detection}

We first merged all the observations to produce   mosaic 
images (with a binsize of 15$''$) in the four energy bands, 
using only the inner part of 
the PSPC detector. In fact the pointing directions were 
appropriately defined to continuosly map the region with the
inner part of the detector, that provides the best sensitivity and
angular resolution (adding the overlapping outer regions with
a different Point Spread Function would degrade the image quality).
The corresponding exposure images
were also produced in the four energy bands.

The final mosaic in the total  energy band, 
corrected both for the different
exposure times and for the vignetting, is displayed in Fig.~\ref{fig:totcor}.
The corresponding (vignetted) exposure map,
used to correct it, is shown in Fig.~\ref{fig:expo}.

We then applied, for each energy band, a source detection algorithm
based  on the following steps:

\begin{itemize}

\item {\em local--detection algorithm}: 
consisting
of a sliding window technique,
where a  detection cell of $3\times3$ pixels is shifted across the  images.
The total counts inside the detection cell
are   compared with a local
background taken from the 16 pixels surrounding the  detection cell itself.
A 3$\sigma$ detection threshold has been applied.
This local detection is used to produce  a preliminary list of sources.

\item {\em production of a smoothed background image}: the source 
list produced in the previous
step is used to remove from the image circular regions 
around each source position.
The resulting  source-free image
is fitted by a two--dimensional spline function to fill the holes and to
 produce a background image.
A background image is produced for each energy band.
 
\item  {\em map--detection algorithm}: 
a new sliding window search is performed,
this time using the background image to extract 
information about the local background.
In other words, for each detection cell, the corresponding background is
extracted, not from a frame surrounding it, but from the background map
at that position.
A second list of sources is produced, again with
a 3$\sigma$ detection threshold.

\item  {\em merging of the source lists}: the two
lists of sources resulting from the two
sliding window searches are merged together, removing duplicates
in order to have a unique list of sources (in each energy band).
Duplicates are removed  
checking  whether the distance between 
two sources is less than twice the sum
of the size of the detection windows   
or less than the FWHM of the point spread
function of the instrument.

\item  {\em  maximum-likelihood method}: a maximum-likelihood method
(Cruddace et al. 1987)
is then applied to the photon lists, using the merged list
of sources as   candidates.
This method considers the ROSAT PSPC point spread function and the
position of the source inside the detector and derives
 a source position and
an existence likelihood.
 Only a detection likelihood larger than 10
(corresponding to a probability of a chance detection smaller than e$^{-10}$)
was considered as a true source.
A final list  of sources with their  positions and
positional uncertainties  is thus produced.

\end{itemize}

This process yielded four lists of sources (one for each
energy band).

The same procedure was also applied on the   individual observations,
this time also considering the external part of the PSPC detector,
resulting in 4$\times$43 lists of sources
(4 energy bands and 43 observations).
This second search was motivated by the fact that, 
e.g. due to source variability,
some sources might have been missed in the previous global analysis.

Finally, all the lists of detected sources  were
cross correlated in  order to
clean the catalogue, removing all the sources with a multiple detection.
By multiple detection we mean two or more sources whose position is
compatible with each other, in which case only the source with higher
existence likelihood has been kept.
This procedure yielded a  final catalogue containing 107 sources.

The count rates and upper limits (2$\sigma$) in each energy
band have been extracted from the photon events tables
at the position of the detected sources.
Two   softness ratios have been also derived:  S/H and M/H,
where H, M and S are the net source counts in the hard, medium and soft
energy bands defined in Sect.~\ref{sect:obs}.

\begin{figure*}[!ht]
\vskip -0.5truecm
\centerline{\psfig{figure=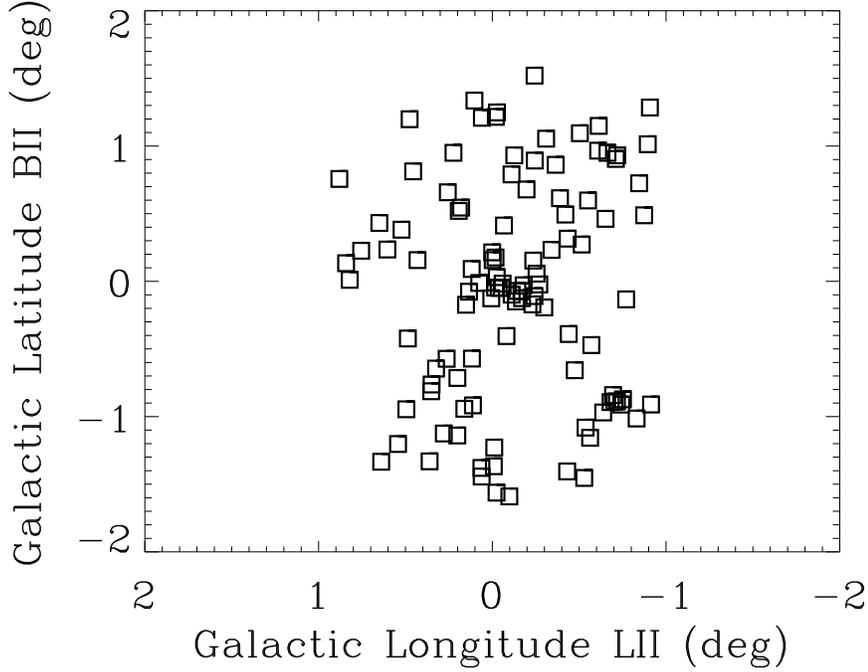,height=120mm}}
\vskip -1.0truecm
\caption{Distribution of the ROSAT sources detected in the galactic center region}
\label{fig:sources}
\end{figure*}

\section{The catalogue of galactic center sources}

\begin{figure*}
\centerline{\psfig{figure=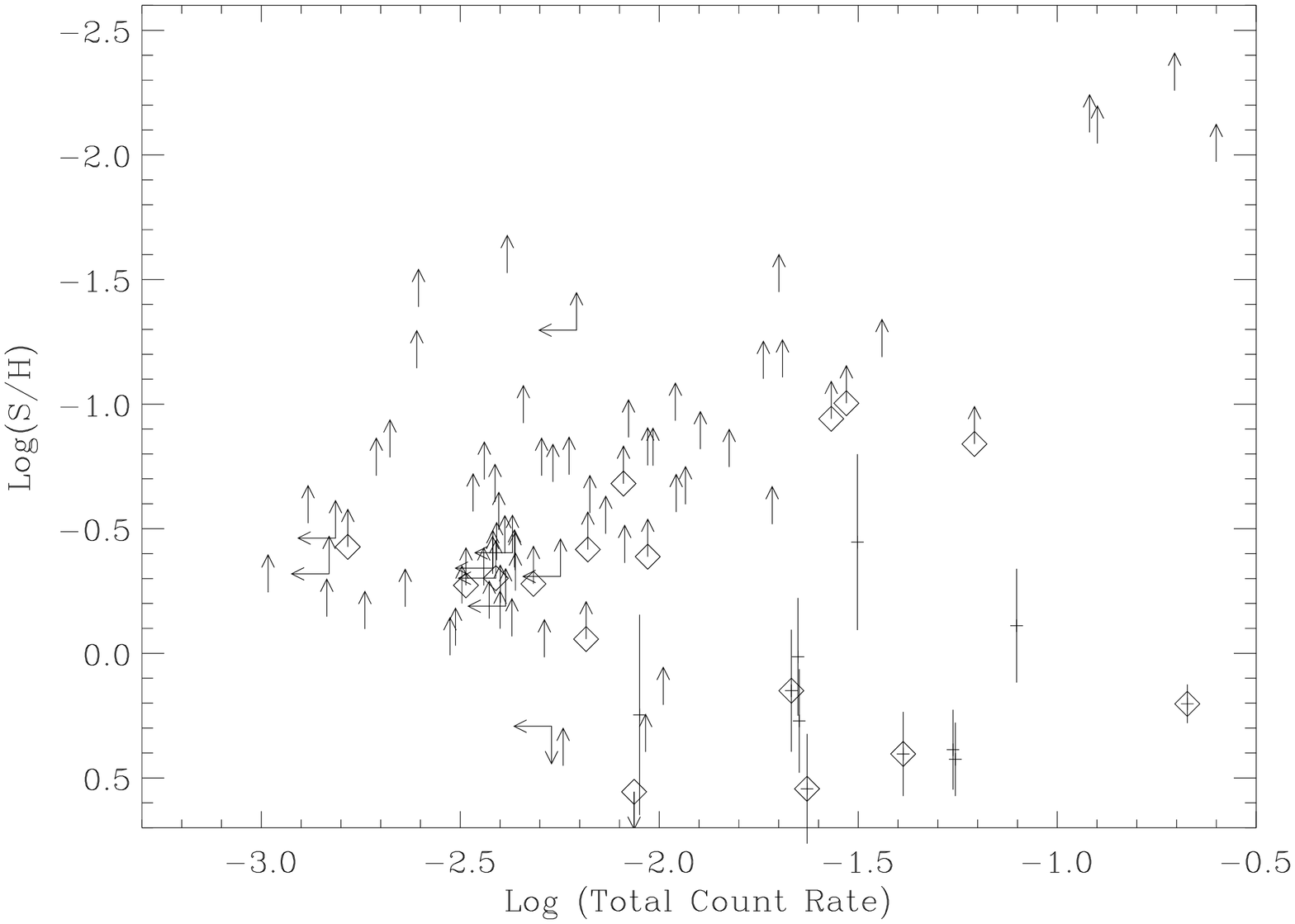,height=150mm,angle=-270}}
\vspace{0cm}
 \caption[]{Softness ratio S/H as a function of the source 
count rate in the
total energy band (0.1-2.4 keV). 
The upper limits on the total count rate refer
to the sources that were detected only in the Soft or in the Hard 
energy bands.
All the error bars are at 1$\sigma$ and the upper/lower limits at
2$\sigma$. The square symbols mark the sources possibly identified with
stars
}
   \label{fig:hr1}
\end{figure*}

\begin{figure*}
\centerline{\psfig{figure=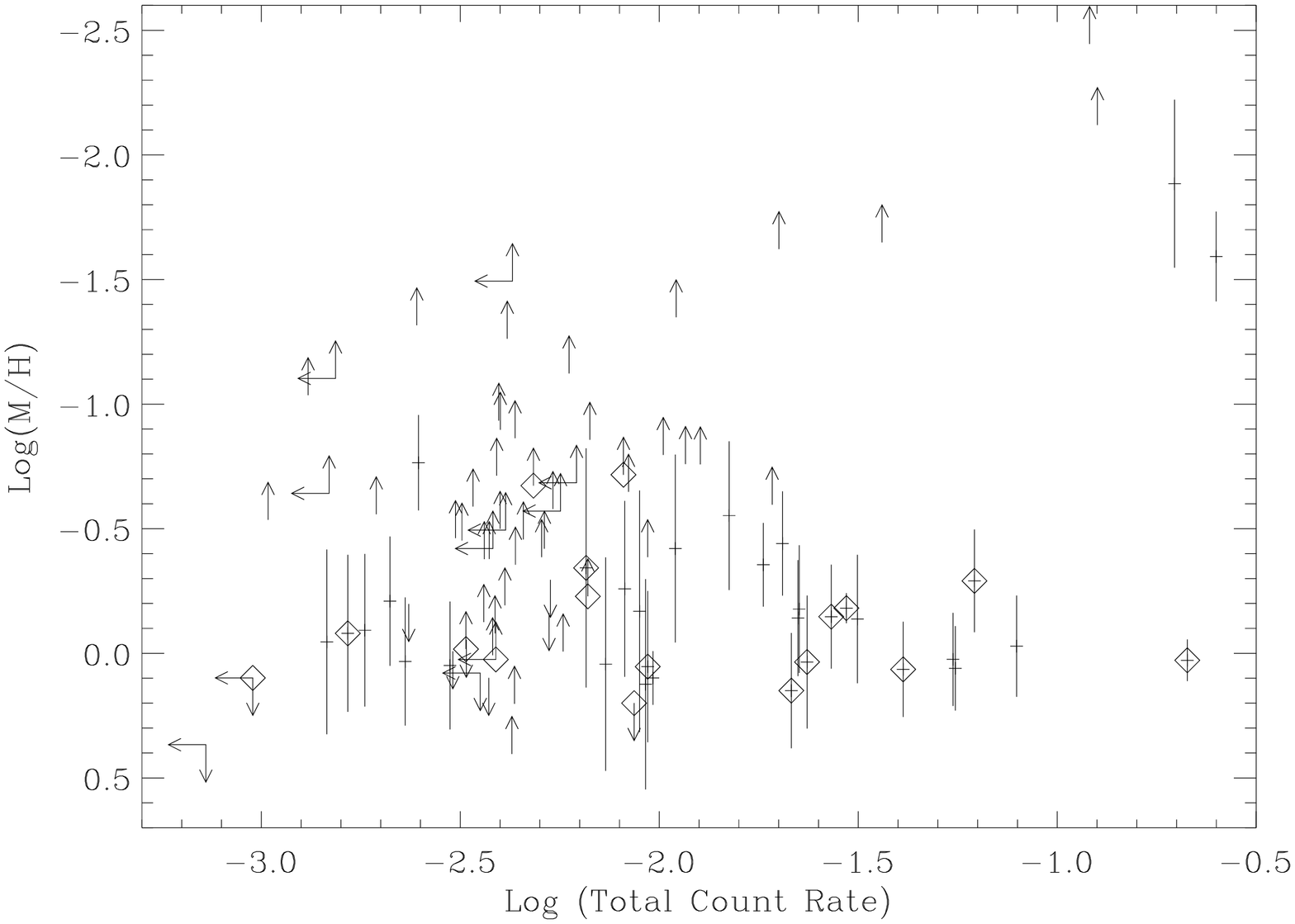,height=150mm,angle=-270}}
\vspace{0cm}
 \caption[]{Softness ratio M/H as a function of the 
source count rate in the
total   energy band (0.1-2.4 keV). The upper limits on the 
total count rate refer
to the sources that were detected only in the Medium or in the Hard 
energy bands.
All the error bars are at 1$\sigma$ and the upper/lower limits at
2$\sigma$. The square symbols mark the sources possibly identified with
stars
}
   \label{fig:hr2}
\end{figure*}

The spatial distribution of the sources detected in the surveyed region
is displayed in Fig.~\ref{fig:sources}. 
The remarkable symmetry of this distribution, and in 
particular the rather uniform source density as a function of
galactic latitude, probably indicates that many sources are 
at a distance much closer than that of the galactic center.
The final catalogue is reported in Table~\ref{catrosat}, where for each source
the identification number, coordinates (J2000), count rate
in the total energy band (0.1--2.4 keV),  softness ratios S/H and M/H and
possible identifications  are reported.
It is possible that a source has an upper limit
in the total energy band, but is detected in one of the single
energy bands.
This is due to the energy-dependent background:
faint sources detected, for example, in the hard energy range,
could have been missed in the total   energy band
 due to a higher background level.
For these sources, the count rate in the  energy band
 where a detection has been found
is shown (and marked).
On the other side, a few sources have been detected in the total
energy band, but have only upper limits in all the other energy ranges;
in this case the softness ratios are missing from Table~\ref{catrosat}.
In Figs.~\ref{fig:hr1} and \ref{fig:hr2} the two softness ratios 
versus the count rate
in the total energy band are shown.

A ROSAT PSPC count rate of 0.01 counts s$^{-1}$
corresponds to about  5$\times10^{-11}$ ergs~cm$^{-2}$~s$^{-1}$
(unabsorbed flux),
assuming a 5 keV bremsstrahlung
spectrum and a column density of 6$\times10^{22}$~cm$^{-2}$.

\section{Discussion}
\label{sect:discussion}

The application of the maximum--likelihood method  to 43 pointings of the
GC region performed with the ROSAT PSPC instrument in 1992--1993 has led to
a new catalogue of soft X--ray sources.
Our catalogue contains 107 detections,
down to a count rate of $\sim$0.001 counts s$^{-1}$ in
the energy range 0.1--2.4 keV.

A correlation with the SIMBAD database resulted in probable identifications
based  on positional coincidence.
The most plausible identifications (objects
falling inside the PSPC error circle) 
are reported in Table~\ref{catrosat}.
Other candidate counterparts
(IRAS sources, for example)
whose positions fall outside the PSPC error are also listed,
when  their large error boxes overlap with the ROSAT error box.
For the possible counterparts listed in column (8) we give in
parenthesis the distance between the optical and X--ray positions, the
spectral type and magnitudes for stars, and  other relevant informations as
described in the notes to Table~\ref{catrosat}.

\vspace{0cm}
\begin{table}[!ht]
\begin{center}
  \caption{X--ray sources not detected}
\label{xrb}
    \begin{tabular}[c]{ll}
\hline
Source & Reference \\

\hline

1E 1743.1--2843 	 & Cremonesi et al. 1999 \\
XTE~J1748--288 		 & In 't Zand et al. 1998 \\
KS~1741--293 		 & In 't Zand et al. 1990 \\
GX+1.1--1.0 		 & Proctor et al. 1978 \\
GX+0.2--1.2  		 & Proctor et al. 1978 \\
GRS~1741.9--2853  	 & Sunyaev et al. 1991 \\
GRS~1734--29		 & Sunyaev et al. 1991b \\
GRS~1743--290 		 & Cordier et al. 1993 \\
GRS~1747--312 		 & Pavlinski et al. 1994 \\
GRS~1747--341 		 & Cordier et al. 1993 \\
1E~1742.5--2845  	 & Watson et al. 1981 \\
1E~1742.7--2902  	 & Watson et al. 1981 \\
1E~1742.9--2849   	 & Watson et al. 1981 \\
1E~1743.1--2852  	 & Watson et al. 1981 \\
1E~1741.2--2859  	 & Mitsuda et al. 1990 \\
GC~X--2  		 & Cruddace et al. 1978 \\
GC~X--4  		 & Cruddace et al. 1978 \\
XTE~J1755--324 		 & Remillard et al. 1997 \\
GRO~J1744--28  		 & Lewin et al. 1996 \\
SAX~J1750.8--2900  	 & Bazzano et al. 1997 \\
1E~1740.7--2942  	 & Mirabel et al. 1992 \\
SAX~J1747.0--2853 = GX~0.2--0.2    &  In 't Zand et al. 1998b \\
XTE~J1739--285 		 & Markwardt et al. 1999 \\
GRS~1739--278  		 & Paul et al. 1996 \\

\hline
\end{tabular}
\end{center}
\end{table}
 

\subsection{X--ray binaries}

Our discovery of ROSAT counterparts to a few previously known X--ray sources,
allows us to significantly improve their  positions.
This is the case of the recently discovered
729~sec
X--ray pulsar AXJ 1740.2--2848 (Sakano \& Koyama 2000, later named 
AX~J1740.1--2847 by Sakano et al. 2000) and of the
other two ASCA sources AX~J1744.3--2940 (Sakano et al. 1999a) and
AX~J1740.3-2904 (Sakano et al. 1999b).

We also inspected the
error boxes of the X--ray binaries indicated in Table~\ref{xrb},
without finding any
ROSAT counterparts in our
catalogue, within a radius of 1$'$ of their position.

\subsection{Foreground Stars}
\label{sec:stars}

\begin{figure}
\centerline{\psfig{figure=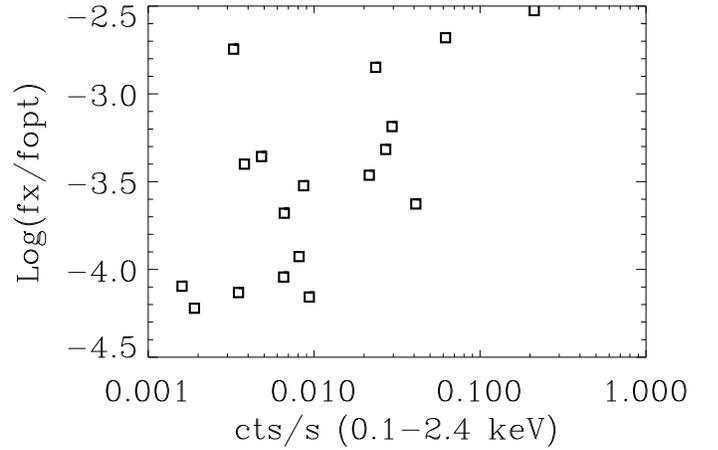,height=70mm,angle=0}}
\vspace{0cm}
 \caption[]{Log(f$_{\rm X}$ / f$_{\rm opt}$) 
as a function of the source count rate in the
total  energy band (0.1-2.4 keV) for the sources identified with stars
}
   \label{fig:stars}
\end{figure}

A number of ROSAT sources have stellar counterparts in their
error circles (see Table~\ref{catrosat}).
To test these associations, we computed
the log(f$_{\rm X}$/f$_{\rm opt}$) as in Voges et al. (1999),
using a  constant conversion factor
of 1.48$\times10^{-11}$ ergs~cm$^{-2}$~cts$^{-1}$,
appropriate for stellar sources (Fleming et al. 1995).
The derived optical to X--ray flux ratios are displayed 
in Fig.~\ref{fig:stars}
as a function of the count rate in the total energy band.

Stars  usually have    log(f$_{\rm X}$/f$_{\rm opt}$) values
in the range [$-$6,$-$1],
depending on the   spectral type.
Therefore, on the basis of the ratio between their X-ray to optical
flux, all these sources could indeed be
stellar X--ray sources.

From Figs.~\ref{fig:hr1} and \ref{fig:hr2} it can be seen that the 
sources tentatively identified with stars have relatively 
softer (or less absorbed) spectra.

\subsection{Supernova Remnants}
\label{snr}

We searched for the possible association of ROSAT sources
with galactic supernova remnants.

The inspection of the Green catalogue of all the SNRs (Green 1998)
located in the region shows that
no ROSAT point sources in our catalogue
 fall inside the region covered by the following
supernova remnants: G0.3+0.0, G0.9+0.1, G1.0--0.1 and
G359.1--0.5.
 
Two ROSAT sources (n~81 and n~71) fall 
on or nearby the northern part of the
radio shell of G359.0--0.9
(23$'$ size).
Source n~3 is located at the center
of the shell like SNR 359.1+0.9 (11$'$$\times$12$'$ size).
This source  can also be associated with the star HD316072
(see Sect.~\ref{sec:stars}).
Diffuse X--ray emission has been recently detected with ASCA from the
G359.0--0.9 shell  and, marginally,
from  359.1+0.9 (Sakano et al. 1999).

\section{Conclusions}

The spatial analysis,   using a maximum-likelihood method,
applied to  public ROSAT PSPC data of the GC region
led to the detection of 107 point-like sources, 
down to a PSPC count rate
of 0.001 counts s$^{-1}$ in the 0.1--2.4 keV energy range.

Using the SIMBAD database, 
42 sources have been associated with objects
at other wavelengths; 20 are probably stars and 5
have been identified with previously known LMXBs. 
Other  sources have been already classified as X--ray objects,
but their nature is still uncertain.
Two sources in our catalogue  could be associated
with   shells of supernova remnants
and one with a molecular cloud or a maser source (n~72).

An error circle of 1$'$ radius centered
on the positions of 25 LMXBs known  
to lie inside the region surveyed with PSPC has been inspected,
leading to negative results.

All the  sources brighter than 0.060 counts s$^{-1}$ 
have been positionally associated  with known objects.

The majority of our sources still remains unidentified. 
They are mostly undetected in the soft and 
medium PSPC energy ranges, they are quite hard and/or severely
absorbed. 
Their  average count rate in the total energy band
(0.01~counts~s$^{-1}$) translates into a luminosity
in the range $10^{35}$--$10^{36}$~erg~s$^{-1}$ (assuming a 
distance of 8.5 kpc) which is suggestive of
mass-transfer from a companion star onto a compact object
in a binary system.

We propose that these still unidentified sources 
belong to a large  population
of X-ray binaries located in the GC region, accreting 
 at low accretion rates, and still largely unknown.
The transient sources recently 
 discovered with ASM--XTE and WFC--BeppoSAX
 would be only the high luminosity tail of this population
 of LMXBs.

\begin{acknowledgements}
We have made use of the ROSAT Data Archive of the Max Planck
Institut f\"ur extraterrestrische Physik at Garching; and of the SIMBAD
database operated at Centre de Donn\'ees astronomiques in Strasbourg.
We thank A.~Parmar for  reading  this 
manuscript and providing helpful comments. 
L.Sidoli acknowledges an ESA Fellowship.

\end{acknowledgements}

\onecolumn

\clearpage

\begin{landscape}

\begin{table*}[htbp]
\begin{center}
  \caption{The ROSAT PSPC Catalogue. 
  The numbers in parenthesis in the first column refer to the sources
detected by Predehl \& Tr\"umper (1994). In column n.5 sources not detected in the total (T)  energy band,
but only in a partial (S,M or H) band, are labelled with a capital letter. 
Column n.8 reports the probable identifications based
on the cross-correlation with SIMBAD database; in parenthesis we indicate the
offset, spectral type, B and V magnitude for stars.
The meaning of the other symbols is: Al*=Eclipsing Binary of Algol type;
V*=variable star; bL*=Eclipsing Binary of beta Lyr type;
*i*= star in double system; Em*=emission line star;
IR=infrared source }
\label{catrosat}
    \begin{tabular}[c]{|l|l|l|c|c|c|c|l|}
\hline 
Source      & R.A.     & Dec.      &  Error &    cts ks$^{-1}$   & Log(S/H)   &   Log(M/H)  & Notes \\
 ID  &    (J2000) & (J2000)  &  ($''$)  &  (0.1--2.4 keV)        &    &  & \\
\hline
 1 & 17 38 26.3  & $-$29 01 47.2   &6.2   &   79  $\pm{6.93}$   &   $-0.111\pm{0.228}$ &  $-0.0287\pm{0.203}$    &  RXJ1738.4--2901      \\
 2 & 17 39 08.6      & $-$28 20 32.5 &14.9  &    11.1  $\pm{2.27} $     &     $-$     &      $-$      &        \\
 3 & 17 39 31.1  & $-$29 09 50.3   &8.9   &  61.9  $\pm{5.86}$     &   $<    -0.840 $   &  $-0.291\pm{0.206}$     &  HD316072 (3$''$; K0;B=11.2;V=9.97); at the center of G359.1+0.9      \\
 4 & 17 39 40.9  & $-$28 51 11.3   &10.0   &  6.61 $\pm{1.36}$     &   $<    -0.416 $   &  $<    -0.229 $    &  V846 Oph (1$''$; (Al* V*) A2; B=10.32; V=9.90)   \\
 5 & 17 40 09.2       & $-$28 47 23  &11.4  &  5.06   $\pm{1.17}$     &   $<    -0.713 $   &  $<   -0.386$    &  AXJ 1740.2--2848 (pulsar )     \\
 6 & 17 40 15.9       & $-$29 03 32.1  &10.0  &  19.2    $\pm{2.54}$      &   $<    -0.519  $    &  $<    -0.597   $      & AXJ1740.3-2904       \\
 7 & 17 40 19.9       & $-$29 00 4.90  &9.8 &   4.34   $\pm{1.30}$     &   $<    -0.334 $   &  $<    -0.863  $     &  \\
 8 & 17 40 23.2       & $-$29 03 58.2  &20.9  &  5.73   $\pm{1.52}$     &   $< 0.451 $   &  $<  -0.00757  $      &    \\
 9  & 17 40 24.0     & $-$28 56 50.5   &3.6   &  55.4 $\pm{3.23}$  &    $ 0.425\pm{0.147}$ &     $0.0601\pm{0.169}$ &  1RXSJ174024.6--285700 (18$''$)  \\
10 & 17 40 41.2     & $-$28 08 50.0   &16.0   &  3.73   $\pm{1.2}$     &   $-$       &  $>$    0.0988       &   \\
 11 & 17 40 42.7       & $-$28 18 13.4  &1.4 &     251  $\pm{6.75}$      &    $<  -1.97 $  &  $-1.59\pm{0.180}$      & RXJ1740.7--2818 (11$''$); 1E1737.5--2817 (47$''$)    \\
 12 & 17 40 45.6       & $-$29 16 31.3  &11.9  &  10.2    $\pm{2.15}$      &   $< 0.2067 $    &  $<    -0.796  $     &      \\
 13 & 17 40 46.7       & $-$28 38 48.1  &9.8 &   9.24  $\pm{1.73} $     &   $< 0.395  $   &  $0.124\pm{0.422}$      &      \\
 14 & 17 40 51.4       & $-$28 18 56.0  &14.8  &  4.26   $\pm{1.32}$     &   $<   -0.0684 $  & $<$ 0.404      &  \\
 15 & 17 41 04.4       & $-$28 15 4.3 & 5.0 &   22.3   $\pm{2.14}$     &      0.0140$\pm{0.236}$    &   $-0.141\pm{0.232}$ &  \\
 16 & 17 41 23.7       & $-$28 47 43.0  &9.1 &   8.92   $\pm{1.60}  $    &      $0.247\pm{0.402}$    &  $-0.169\pm{0.484}$  &    \\
 17 & 17 41 33.8       & $-$28 40 35.0  &5.9 &   20.3    $\pm{1.92} $      &  $< -1.107   $    &  $-0.441\pm{0.209}$     &      \\
 18 & 17 41 36.7     & $-$29 25 30.6   &22.5   &    4.32     $\pm{1.22}$    &    $<   -0.345 $      & $<$ 0.202     &  \\
 19 & 17 41 41.8       & $-$28 33 22.1  &6.4 &   22.4    $\pm{2.08} $      &       $0.271\pm{0.208}$    &   $-0.177\pm{0.256}$    &       \\
 20  & 17 41 58.0     & $-$29 05 34.5   &5.0   &  21.5 $\pm{1.97}$  &    $0.15\pm{0.24}$    &     $0.150\pm{0.231}$ &  HD160572 (2$''$; F3V ;B=9.59;V=9.16)      \\
21 & 17 42 06.5     & $-$27 54 12.5   &18.7   &  3.07   $\pm{0.9} $     &   $-$       &        $-$       &   \\
 22  & 17 42 15.2     & $-$29 14 59.2   &3.2   &  54.70 $\pm{3.43}$  &     $0.387\pm{0.16}$  &     $0.0245\pm{0.187}$ &       \\
 23 & 17 42 16.9       & $-$28 37 1.2 & 16.9  &  3.99   $\pm{1.00}$     &   $<    -0.203 $  &   $<$    $-0.897 $      &  \\
 24 & 17 42 17.5       & $-$28 56 47.6  &11.6  &  3.98   $\pm{1.11}$     &   $<   -0.099 $   &  $<    -0.499  $    &        \\
 25 & 17 42 27.8   &  $-$28 14 55.9  &13.0  &   2.1 $\pm{0.7}^{(H)}$     &       $<   -0.342 $        &    $<$    $-0.421$  & \\
 26 & 17 42 30.3       & $-$28 44 56.3  &1.8 &     212   $\pm{6.65} $     &      $0.203\pm{0.077}$     &  $0.0280\pm{0.0836}$    &  V2384 Oph (1$''$; (bL* V*) G3/G5V; B=9.75; V=9.02)     \\
 27 & 17 42 41.3       & $-$29 02 13.6  &8.7 &   11.6    $\pm{1.45}$      &   $<    -0.599  $    &  $<    -0.760  $       &      \\
 28  & 17 43 19.5     & $-$29 14 1.10   &15.7  & 5.40   $\pm{1.14}$ &    $< -0.688 $     &    $<  -0.579 $ &  \\
 29 & 17 43 21.0     & $-$29 08 29.9   &19.2   &  3.1   $\pm{0.99}$     &   $<  -0.0307   $     &   $< -0.462 $      &  \\
 30 & 17 43 32.7   & $-$28 07 25.0 &5.5   &   23.5     $\pm{1.9}$      &   $0.544\pm{0.220}$     & $0.0348\pm{0.267}$  &  HD316199  (2$''$; K5; B=10.6)    \\
 31 & 17 43 40.2   &  $-$28 22 27.7  &52.0  &     1.9  $\pm{0.7}^{(H)}$   &      $<    -0.302 $        &    $<$    0.0246  &  HD316212  (47$''$; K5; B=11.3; V=9.9)  \\
 32 & 17 43 51.1       & $-$28 46 43.7  &11.6  &  11.0    $\pm{1.94}$      &   $<    -0.934 $  &  $-0.420\pm{0.377}$     &      \\
 33 & 17 43 53.8       & $-$29 06 21.9  &15.1  &  2.7    $\pm{1.0}^{(H)}$    & $<   -0.308 $  &  $<  -0.571$     & IRAS17407--2904 (44$''$)      \\
 34 & 17 43 55.6   &  $-$28 29 57.9  &14.1  &  4.09   $\pm{1.23}$     &    $<   -0.402 $      &  $<$ $-0.192$      &  IR [OF84] 18 (37$''$); IRAS 17407--2829 (49$''$) \\
 35 & 17 44 03.1   &    $-$28 30 1.5   & 15.3   &  5.33  $\pm{1.4}$    &    $-$       &  $>$ $-0.295$      &  \\
 36  & 17 44 17.6     & $-$29 39 48.0   &6.6   & 14.9 $\pm{1.79}$    &    $< -0.748 $     &   $-0.553\pm{0.298}$  &  AXJ1744.3--2940  (19$''$)    \\
\hline
\end{tabular}
\end{center}
\begin{small}
\end{small}
\end{table*}


\clearpage

\addtocounter{table}{-1}

\begin{table*}[htbp]
\begin{center}
  \caption{Continued}
\label{catrosat}
    \begin{tabular}[c]{|l|l|l|c|c|c|c|l|}
\hline 
Source      & R.A.     & Dec.      &  Error &    cts ks$^{-1}$   & Log(S/H)   &   Log(M/H)  & Notes   \\
 ID  &    (J2000) & (J2000)  &  ($''$)  &  (0.1--2.4 keV)        &    &  & \\
\hline
 37 & 17 44 27.4    & $-$29 03 29.7   &8.4    & 2.9 $\pm{0.356}  $ &     $<  0.00717  $    & 0.0489$\pm{0.256}$     &       \\
38 & 17 44 46.2     & $-$27 47 27.3   &17.4   &  5.3  $\pm{2.0} $      &   $-$       &  $>    -0.161 $       &        \\
 39 & 17 44 47.0     & $-$28 49 28.3   &17.3   &    1.27     $\pm{0.316}$   &   $-$      &       $-$       &    \\
 40  & 17 44 47.3     & $-$29 07 27.9   &13.9  & 2.3 $\pm{0.355}$   &       $-     $  &  $>   -0.198$    &   \\
 41 & 17 44 53.3     & $-$28 51 39.3   &10.0   &   2.8   $\pm{0.4}^{(H)}$    &      $< -0.404 $        &     $<   -1.494$ & IR MGM 1--3 (6$''$)  \\
  42 (4) & 17 45 00.1    & $-$28 51 24.9   &15.0   & 3.9 $\pm{0.436} $ &     $<   -0.495  $    & $<   -0.934 $   &  1E 1741.7--2850   (56$''$)  \\
 43  & 17 45 03.8     & $-$29 10 47.8   &16.2  & 1.2 $\pm{0.279}$   &       $-$   &  $-$      &  IR GCS13 (18$''$)     \\
 44 (13) & 17 45 18.1     & $-$29 06 21.4   &10.5  & 1.3 $\pm{0.243} $  &      $<    -0.522  $ &   $<  -1.036  $   &  \\
 45 & 17 45 22.5   &  $-$28 17 32.5  &15.1  &  3.62   $\pm{1.24}$     &   $<$$-0.273 $      &   $< -0.126 $    &  \\
 46  & 17 45 26.1     & $-$28 56 32.6   &10.9  & 0.38 $\pm{0.10}^{(M)}$    &   $-$      &  $>$    0.0986       &  CSI-28-17423 (7$''$; B; B=11.5;V=11.2); HD316223 (7$''$; G;B=1.3)      \\
 47 & 17 45 28.0     & $-$29 12 4.5   & 26.2   &   1.2  $\pm{0.3}^{(H)}$ &        $<    -0.462 $   &     $<    -1.103 $    & IRAS 17422--2911   (51$''$)    \\
 48 & 17 45 29.0   &  $-$28 09 25.1  &10.3  &  8.37  $\pm{1.60}$    &   $<$$-0.866$      &   $< -0.648 $      &        \\
 49 (12) & 17 45 30.3     & $-$29 07 06.8   &2.3   &  9.6 $\pm{0.499} $  &      $<    -0.753  $ &  $0.0989\pm{0.108}$   &  \\
 50 (8)  & 17 45 32.5     & $-$28 59 47.6   &25.6  & $<$0.30   &     $-$         &     $-$   &   \\
  51 & 17 45 32.6    & $-$28 47 17.3   &11.4   & 1.4 $\pm{0.275} $ &     $<   -0.147  $    & $-0.0456\pm{0.371}$  &      \\
 52 (11) & 17 45 33.1     & $-$29 05 50.0   &8.0   &  1.0 $\pm{0.220}$  &    $< -0.2446$     &   $<$$-0.536$        &  \\
 53  & 17 45 39.4     & $-$29 17 33.6   &16.4  & 0.9 $\pm{0.2}^{(H)}$       &    $<  -0.3189 $    &    $<-0.642 $     &    \\
 54 (7) & 17 45 40.7  & $-$29 00 29.4   &13.7   &   2.46 $\pm{0.48}$ &       $<    -1.145 $  &     $<    -1.316 $   & SgrA*   (9$''$)  \\
 55  & 17 45 41.7     & $-$29 08 52.7   &17.4  & 0.33 $\pm{0.10}^{(M)}$    &   $-$      &  $>$     0.367       &        \\
  56 & 17 45 43.1    & $-$28 59 36.7   &18.7   & 5.5 $\pm{0.5}$ &     $<    -1.297 $    &  $<   -0.684  $  &        \\
 57 (14) & 17 45 44.0     & $-$29 13 22.9   &11.2  &  1.6 $\pm{0.289}$   &       $<   -0.427   $  & $-0.0802\pm{0.315}$    &   HD316232  (6$''$; O+...; B=11.1 , V=10.4 )   \\
 58 (9) & 17 45 44.1    & $-$29 04 59.4   &5.5    & 2.2 $\pm{0.287}   $  &      $<   -0.1869  $ & 0.0329$\pm{0.257}$   &    \\
  59 (6) & 17 45 45.6    & $-$28 58 29.2   &4.8    & 2.4 $\pm{0.286}$ &     $<    -1.39  $    &  $-0.765\pm{0.191}$  &     \\
  60 (3) & 17 45 50.7    & $-$28 52 43.7   &5.9    & 2.1 $\pm{0.275} $ &     $<   -0.786  $    &  $-0.209\pm{0.259}$  &   \\
 61 (10) & 17 45 52.5     & $-$29 07 49.4   &6.6   &  1.8 $\pm{0.266}$  &      $<  -0.0979   $ &  $-0.0926\pm{0.306}$   &     \\
 62 & 17 46 01.8   & $-$28 29 16.7 &16.4 &  8.64    $\pm{1.5} $     &   $>$    0.556      & $>$ 0.199       &  HD316297 (11$''$; K7; B=11.4; V=10.0)  \\
 63  & 17 46 05.5     & $-$29 30 54.8   &3.2   &  121 $\pm{4.37}$ &     $<   -2.091$      &    $<$     -2.446  &  A1742--294 (13'')   \\
 64  & 17 46 06.2     & $-$29 40 9.40   &13.0  & 3.27   $\pm{0.80}$ &    $< -0.272$       &   $<  -0.0168  $       &   V734 Sgr (4$''$; (Al* V*?); B=13.00)  \\
 65 (5) & 17 46 07.3    & $-$28 59 50.2   &5.7    & 1.9 $\pm{0.271}$ &     $<   -0.713  $    &  $<   -0.558 $   &    \\
 66 & 17 46 08.2   &  $-$28 17 56.4  &12.5  &  5.12   $\pm{1.27}$     &    $-$      &       $-$     &  \\
  67 (2) & 17 46 14.1    & $-$28 51 44.9   &18.2   & 4.1 $\pm{0.623} $ &     $<    -1.53  $    &  $< -1.26 $   &  1E 1742.9--2849   (30$''$) \\
 68 & 17 46 31.6   &  $-$28 10 29.0  &13.5  &  3.90   $\pm{1.15}$     &   $<$$-0.374 $      &   $< -0.713 $      &     \\
  69 (1) & 17 46 39.2    & $-$28 53 52.4   &1.8    & 29.5 $\pm{0.864}$ &     $<    -1.00  $    &  $-0.181\pm{0.060}$  &  HD316314 (0$''$; F0; B=9.94; V=9.51); 1E1743.4--2852 (2$''$)   \\
 70 & 17 47 00.4       & $-$29 13 1.90  &8.2 &   18.2    $\pm{1.42}$      &   $< -1.102 $  &  $-0.355\pm{0.167}$     &      \\
 71  & 17 47 00.8     & $-$30 11 27.1   &25.6  & 8.8 $\pm{2.54}$ &     $-$         &     $-$   &  inside the shell of the SNR G359.0--0.9     \\
 72 & 17 47 03.7       & $-$29 41 2.70  &17.1  &  3.27   $\pm{1.05}$     &        $-$    &  $>$  $-0.0539$       &  OH359.5--0.7(13$''$; Molecular Cloud); [TVH89]203(15$''$; Maser)    \\
\hline
\end{tabular}
\end{center}

\end{table*}


\clearpage

\addtocounter{table}{-1}

\begin{table*}[htbp]
\begin{center}
  \caption{Continued}
\label{catrosat}
    \begin{tabular}[c]{|l|l|l|c|c|c|c|l|}
\hline 
Source      & R.A.     & Dec.      &  Error &    cts ks$^{-1}$   & Log(S/H)   &   Log(M/H)  & Notes   \\
 ID  &    (J2000) & (J2000)  &  ($''$)  &  (0.1--2.4 keV)        &    &  & \\
\hline

 73 & 17 47 05.3   &  $-$28 08 54.8  &6.4  &   27.0   $\pm{2.81}$      &    $<   -0.941 $      &  $-0.147\pm{0.208}$   &  BN Sgr (5$''$; (Al* V*) F3V; B=9.60; V=9.28)   \\
 74 & 17 47 14.9     & $-$30 01 57.5   &17.7   &  3.6   $\pm{1.2}$     &    $<-0.697 $      &  $<  -0.378 $       &  \\
 75  & 17 47 15.5     & $-$29 58 04.3   & 8.0  & 20 $\pm{2.48 }$ &     $<   -1.450 $     &    $<$    $ -1.62$  &  G359.23--0.92 (The Mouse)   \\
 76  & 17 47 23.5     & $-$30 00 39.3   &22.4  & 36.2 $\pm{3.61}$ &     $<   -1.189 $     &   $<$    $-1.649$  &   \\
77 & 17 47 25.4     & $-$30 02 40.7   &3.6    &  126    $\pm{6.0}$       &      $<     -2.046 $ &    $<    -2.12 $   &  SLX1744--300   (27$''$)  \\
 78  & 17 47 25.9     & $-$29 59 57.8   &2.7   &  197.2   $\pm{7.30}$ &     $<    -2.258 $    &  $-1.88\pm{0.337}$  &  SLX1744--299 (15$''$)     \\
 79  & 17 47 29.9     & $-$29 58 57.7   &17.3  & 11.0 $\pm{2.09}$ &     $<  -0.567 $      &    $<$     $-1.348   $  &   \\
 80 & 17 47 31.0   &  $-$28 13 46.8  &5.5  &   31.4   $\pm{3.64}$      &   $-0.446\pm{0.353}$     &  $-0.138\pm{0.258}$   &      \\
 81  & 17 47 37.3     & $-$30 10 23.9   &27.6  & 12.6 $\pm{3.12}$ &     $<  -0.8195  $    &   $< -0.758 $  &  near the shell of the SNR G359.0--0.9  \\
 82 & 17 47 54.2     & $-$29 59 8.4   & 15.3   &  4.1   $\pm{1.25}$     &    $-$      &     $-$        &        \\
 83 & 17 48 07.8       & $-$29 07 58.8  &13.9  &  4.82   $\pm{1.13}$     &    $< -0.279 $  &   $<$    $-0.673$      & CD--29 14004 (7$''$; B+...; B=11.83; V=11.05)  \\
 84 & 17 48 25.1   & $-$28 44 24.5 &15.7 &  3.19    $\pm{0.84} $     &   $<   -0.200 $    &  $<   -0.452$      &    \\
 85 & 17 48 28.7   &   $-$29 00 33.0   &12.8   &  3.81  $\pm{0.96}$     &    $< -0.319 $     &  $<   0.00770 $       &       \\
 86  & 17 48 35.3     & $-$29 57 23.3   &10.7  & 8.11 $\pm{1.36}$    &    $< -0.681$       &    $<-0.717 $  &  HD316341  (6$''$; (Em*) O+...;B=9.54;V=9.06)    \\
  87 & 17 48 49.4    & $-$30 01 03.9   &10.7   & 5.1 $\pm{1.02}$ &     $<   0.0156 $    & $<   -0.420 $  &     \\
 88 & 17 48 53.4       & $-$29 08 6.70  &7.5 &   8.19   $\pm{1.27}$     &    $< -0.363 $  &  $-0.259\pm{0.353}$     &      \\
 89 & 17 48 54.4   & $-$28 59 39.2 &11.0 &  6.55   $\pm{1.31}$     &   $<  -0.0571  $     &  $-0.343\pm{0.480}$  & HD316308 (9$''$; K0; V=9.0)   \\
 90 & 17 49 25.8   &  $-$29 01 57.2  &16.7  &  1.9    $\pm{0.7}^{(H)}$   & $<    -0.189 $  &  $<  -0.494 $      &  \\
 91 & 17 49 28.6       & $-$29 18 58.8  &4.10  &  41.0   $\pm{2.7} $     &  $0.404\pm{0.169}$   & $0.0645\pm{0.191}$      &  HD161907 (6$''$; (*i*); B=8.33;V=8.05);       \\
    &           &   &  &       &        &          & CD--29 14038B (9$''$; (*i*);V=13.0)     \\
 92 & 17 49 37.8   & $-$29 03 24.1 &10.5 &  9.35    $\pm{1.54}$     &  $<    -0.754 $   &   $<$  $-0.386$     &     \\
 93 & 17 49 41.4     & $-$29 17 12.7   &17.3   &  3.5   $\pm{1.02}$     &    $-$      &     $-$        &  HD316418  (16$''$; F0; B=9.94; V=9.46)  \\
 94  & 17 50 04.9       & $-$30 08 29.3  &16.4  &  7.33   $\pm{1.61}$     &   $<    -0.480 $  &  $0.0434\pm{0.428}$     &      \\
 95 & 17 50 07.3       & $-$30 01 55.2  &14.1  &  3.73   $\pm{1.17}$     &   $<    -0.139 $  &   $<    -0.378 $    &    \\
 96 & 17 50 24.9     & $-$29 34 53.5   &16.4   &  1.5  $\pm{0.50}^{(M)}$  &        $-$    &   $>$    0.0794    &      \\
 97 & 17 50 29.4   &  $-$29 00 7.2 & 10.3  &  5.93   $\pm{1.22}$     &  $<    -0.717 $   &   $<$  $-1.123$     &  1RXPJ175029.8--285957 (10$''$)  \\
98 & 17 50 33.8     & $-$29 21 11.4   &15.3   &  2.6  $\pm{0.8} $     &   $-$       &  $-$        &   \\
 99 & 17 50 41.3    &  $-$29 16 44.5  &7.5 &   9.35   $\pm{1.38}$     &    $< -0.388 $  & $0.0536\pm{0.304}$      & HD162120 (3$''$; (*i*)  A2V; B=8.51; V=8.33)     \\
100 & 17 50 58.0     & $-$29 39 00.1   &13.3   &  3.24   $\pm{1.05}$     &    $-$      &     $-$        &   \\
 101 & 17 51 11.1     & $-$29 35 41.9   &9.8    &  3.41   $\pm{0.99}$     &    $< -0.570 $     &   $<  -0.589  $       &  \\
 102 & 17 51 25.9     & $-$29 37 44.4   &11.0   &  4.56   $\pm{1.17}$     &    $< -0.924 $     &   $<   -0.457 $       &  \\
 103 & 17 51 36.7     & $-$29 05 31.8   &17.1   &  3.02   $\pm{0.86}$     &    $-$       & $>$    0.0793       &  \\
 104 & 17 51 38.5       & $-$29 50 26.3  &14.8  &  6.69   $\pm{1.93}$     &    $< -0.563  $  &   $<$  $-0.857$     &       \\
 105 & 17 51 41.4   & $-$29 18 49.5 &12.6 &  4.35    $\pm{1.28}$     &   $<    -0.252 $  &  $<$ $-0.356$     &      \\
106 & 17 51 42.2     & $-$29 45 47.4   &31.0   &  3.5  $\pm{1.1}^{(S)}$   &        $>$   0.293    & $-$      &        \\
 107 & 17 52 21.2   &  $-$29 04 32.2  &10.9  &   3.9    $\pm{1.12}$     &    $<   -0.608 $      &  $<$ $-0.0804$     &     \\

\hline
\end{tabular}
\end{center}
\begin{small}
\end{small}
\end{table*}

\end{landscape}

\clearpage


\begin{thebibliography}{}

\bibitem{}
Bazzano A., Heise J., Ubertini P., et al., 1997, IAU Circ. 6597 

\bibitem{}
Bradt H., Levine A.M., Remillard R.A., Smith D.A., 2000,  
In ``X--ray Astronomy 1999; Stellar Endpoints, AGN and the
Diffuse Background", Bologna, Sep. 1999,
Eds. G.Malaguti, G.Palumbo and N.White, pub. Gordon and Breach, in press
(astr-ph/0003438)


\bibitem{}
Cordier B., et al., 1993, In ``The Second Compton Symposium", AIP 304, 
Eds. C.E.Fichtel, N.Gehrels and J.P.Norris (New York), 446

\bibitem{}
Cremonesi D.I., Mereghetti S., Sidoli L., Israel G.L., 1999, A\&A 345, 826

\bibitem{}
Cruddace R.G., Hasinger G.R., Schmitt J.H.M.M. 1987, 
In ``Astronomy from Large Databases", Murtaugh F., Heck A. (eds.), p.177 

\bibitem{}
Cruddace R.G., Fritz G., Shulman S., et al., 1978, ApJ 222, L95

\bibitem{}
Green D.A., 1998, In ``A catalogue of galactic Supernova  Remnants'',
{\sc http://www.mrao.cam.ac.uk/surveys/snrs/}

\bibitem{}
Fleming T.A., Molendi S., Maccacaro T., Wolter A., 1995, ApJ Suppl.Ser. 99, 701


\bibitem{}
In 't Zand J.J., Heise J., Smith M., et al., 1998, IAU Circ. 6840

\bibitem{}
In 't Zand J.J., Bazzano A., Cocchi M., et al., 1998b, IAU Circ. 6846

\bibitem{}
In 't Zand,  et al., 1990, Adv. Space Res. 11, 187

\bibitem{}
Lewin W.H.G., Rutledge R.E., Kommers J.M., et al., 1996, ApJ., 462, L39

\bibitem{}
Markwardt C.B., Marshall F.E., Swank J.H., et al., 1999, IAU Circ. 7300

\bibitem{} 
Mirabel F., Rodriguez L.F., Cordier B., et al., 1992, Nat 358, 215

\bibitem{} 
Mitsuda K., Takeshima T., Kii T., Kawai N., 1990, ApJ 353, 480
 
\bibitem{}
Paul J., Bouchet L.,  Churazov E., Sunyaev R., 1996, IAU Circ.  6348 

\bibitem{}
Pavlinskii M., Grebenev S.A., Sunyaev R.A.,  1994, ApJ 425, 110

\bibitem{}
Pfeffermann E., Briel U.G., Hippmann H., et al. 1986, SPIE 733, 519

\bibitem{}
Predehl P., Tr\"umper J., 1994, A\&A 290, L29
 
\item
Predehl P. et al. 1995, in ``Nuclei of Normal Galaxies", eds. R. Genzel \& A.I. Harris 
(NATO ASI Ser. C, 445),  (Dordrecht:Kluwer), 21 

\bibitem{}
Proctor R.J., Skinner G.K., Willmore A.P.,  1978, MNRAS 185, 745

\bibitem{}
Remillard R., Levine A., Swank J., Strohmayer T., 1997, IAU Circ. 6710

\bibitem{}
Sakano M., Imanishi K., Tsujimoto M., et al., 1999a, ApJ 520, 316
 
\bibitem{}
Sakano M., Yokogawa J., Mukarami H., et al., 1999b, In Proc. of the ``Japanese German 
Workshop on High Energy Astrophysics", eds.  Becker W. \&  Itoh W., MPE Report 270, 113 

\bibitem{}
Sakano M.,  Koyama K.,  2000, IAU Circ. 7364

\bibitem{}
Sakano M., Torii K., Koyama K., et al., 2000, PASJ 52, in press (astro-ph/0008331)

\bibitem{}
Sidoli L., Mereghetti S., Israel G.L., et al.,  1999, ApJ  525, 215

\bibitem{}
Sunyaev R.A., Pavlinskii M., Churazov E., et al., 1991, SovA. Lett. 17, 42

\bibitem{}
Sunyaev R.A., Churazov E., Gilfanov M., et al., 1991b, Adv. Space Res.  11, 177

\bibitem{}
Te Lintel Hekkert P., Versteege-Hansel H.A., Habing H.J., et al.,  1989, A\&AS 78, 399 (TVH89)


\bibitem{}
Ubertini P.,  Bazzano A., Cocchi M., et al., 1999, Proc. ``Third INTEGRAL Workshop", Taormina 1998,
Astroph. Lett. and Communications 38, 301  

\bibitem{}
Voges W., Aschenbach B., Boller Th., et al., 1999, A\&A 349, 389

\bibitem{} 
Watson M.G., Willingale R., Grindlay J.E., et al., 1981, ApJ 250, 142
 
\bibitem{}
Zimmerman H.U., Becker W., Belloni T., et al., 1994, EXSAS User's Guide, MPE Report 257

 
\end{thebibliography}
\end{document}